\documentclass{aa}  
\usepackage{natbib}
\usepackage{graphicx}
\usepackage{txfonts}
\usepackage{hyperref}
\begin{document} 

  \title{A flare in the optical spotted in the changing-look \\ Seyfert NGC 3516}
  \subtitle{}
  \author{D. Ili\'c\inst{1}
     \and
    V. Oknyansky\inst{2}
    \and 
    L. \v C. Popovi\'c\inst{1,3} 
    \and
    S. S. Tsygankov\inst{4,5}
    \and
    A. A. Belinski\inst{2}
    \and 
    A. M. Tatarnikov\inst{2}
    \and
    A. V. Dodin \inst{2}
    \and 
    N. I. Shatsky\inst{2}
    \and 
    N. P. Ikonnikova\inst{2}
    \and 
    N. Raki\'c\inst{1,6}
    \and 
    A. Kova\v cevi\'c\inst{1}
    \and 
    S. Mar\v ceta-Mandi\'c\inst{1,3}
    \and 
    M. A. Burlak\inst{2}
    \and 
    E. O. Mishin\inst{2}
    \and 
    N. V. Metlova\inst{2}
    \and 
    S. A. Potanin\inst{2,7}
    \and
    S. G. Zheltoukhov\inst{2,7}
}
%\fnmsep          \thanks{Just to show the usage of the elements in the author field}

\institute{Department of Astronomy, Faculty of Mathematics, University of Belgrade, Studentski trg 16, 11000 Belgrade, Serbia\\
         \email{dilic@matf.bg.ac.rs}
         \and Sternberg Astronomical Institute, Moscow M.V. Lomonosov State University, Universitetskij pr., 13, Moscow, 119234, Russia
%\email{oknyan@mail.ru, aleks@sai.msu.ru, samsebedodin@gmail.com, ikonnikova@gmail.com, andrew@sai.msu.ru, marina.burlak@gmail.com, yegor1993@gmail.com, nm@msu-crimea.ru, potanin@sai.msu.ru, kolja@sai.msu.ru, sergei.jeltouhov@yandex.ru}
       \and
       Astronmical Observatory Belgrade, Volgina 7, 11000 Belgrade, Serbia
%       \email{lpopovic@aob.rs, sladjana@aob.rs}
\and
Department of Physics and Astronomy, FI-20014 University of  Turku , Finland
\and 
Space Research Institute of the Russian Academy of Sciences,  Profsoyuznaya Str . 84/32, Moscow 117997, Russia
        \and
        Faculty of Natural Sciences and Mathematics, University of Banjaluka, Mladena Stojanovi\'ca 2, 78000 Banjaluka, Republic of Srpska, Bosnia and Herzegovina
%       \email{nemanja.rakic@unibl.rs}
\and
Faculty of Physics, Moscow M.V. Lomonosov State University, Leninskie gory 1, Moscow, 119991, Russia
 }

  \date{Received Jan xx, 2020; accepted Apr 02, 2020}

  \abstract
   {We present observations from the short-term intensive optical campaign (from September 2019 to January 2020) of the changing-look Seyfert NGC 3516. This active galactic nucleus is known to have strong optical variability and has changed its type in the past. It has been in the low-activity state in the optical since 2013, with some rebrightening from the end of 2015 to the beginning of 2016, after which it remained dormant.
  }
   {We aim to study the photometric and spectral variability of NGC 3516 from the new observations in \textit{U}- and \textit{B}-bands and examine the profiles of the optical broad emission lines in order to demonstrate that this object may be entering a new state of activity.   
 }
   {NGC 3516 has been monitored intensively for the past 4 months with an automated telescope in \textit{U} and \textit{B} filters, enabling accurate photometry of 0.01 precision.  Spectral observations were triggered when an increase in brightness was spotted. We support our analysis of past-episodes of violent variability with the UV and X-ray long-term light curves constructed from the archival \textit{Swift}/UVOT and \textit{Swift}/XRT data. }
   {An increase of the photometric magnitude is seen in both \textit{U} and \textit{B} filters to a maximum amplitude of 0.25 mag and 0.11 mag, respectively. During the flare, we observe stronger forbidden high-ionization iron lines ([\ion{Fe}{vii}] and [\ion{Fe}{x}]) than reported before, as well as the complex broad H$\alpha$ and H$\beta$ lines. This is especially seen in H$\alpha,$ which appears to be double-peaked. It seems that a very broad component of $\sim$10,000 km s$^{-1}$ in width in the Balmer lines is appearing. The trends in the optical, UV, and X-ray light curves are similar, with the amplitudes of variability being significantly larger in the case of UV and X-ray bands.
   }
   {The increase of the continuum emission, the variability of the coronal lines, and the very broad component in the Balmer lines may indicate that the AGN of NGC 3516 is finally leaving the low-activity state in which it has been for the last $\sim$3 years.}


\maketitle
%________________________________________________________________
%$-14 \leq \lg \rho / \mathrm{[g\, cm^{-3}]} \leq 0 $,
%   $ 8.8 \leq \lg e / \mathrm{[erg\, g^{-1}]} \leq 17.7$. 

\section{Introduction}

The open question concerning the mechanism by which activity is triggered and evolves in active galactic nuclei (AGNs) is still pertinent, especially because we know that growth of supermassive black holes in the centers of galaxies happens during the AGN stage. We hope to get closer to answering the above question by studying the extremely variable AGNs,  that is, the class of so-called changing-look AGNs (CL AGNs). These objects show extreme changes of emission line intensities, with sometimes almost complete disappearance and reappearance of the broad component \cite[see, e.g.,][etc.]{ly84,kf85,de14,ok19a,ok19b}. Recently, the CL AGNs have come into immediate focus, with large monitoring and spectroscopic surveys \cite[e.g.,][]{ru16,ml16}. There are even speculations that each strongly variable AGN could almost be a CL AGN if constantly observed \cite[see discussion in][]{ok17a}.

There are several physical processes that could cause such dramatic changes. The main ones are: changes of accretion disk structure \cite[e.g.,][]{st18}, broad-line region (BLR) occultation by obscuring material \cite[e.g.,][]{el12}, tidal disruption events (TDEs), supernova explosions, tidal stripping of stars \cite[see, e.g.,][]{wa12, ca15, ko17}, or a combination of the above. So far, it seems that the significant change of the accretion rate is responsible for the change \cite[][]{nd18,sc19}, whereas obscuration is less favored as a major cause \citep{ok17a,ma18}. However monitoring and multiwavelength observations of CL AGNs are needed to discriminate among the different scenarios \cite[e.g.,][]{ml19}.

NGC 3516 is one of the original six Seyfert galaxies \citep{se43}. It is a face-on spiral galaxy at a distance of between $D\approx$38 Mpc and 66 Mpc\footnote{The NASA/IPAC Extragalactic Database (NED) is operated by the Jet Propulsion Laboratory, California Institute of Technology, under contract with the National Aeronautics and Space Administration.}. NGC 3516 has previously been the subject of several optical monitoring campaigns \cite[e.g.,][etc.]{ld93,wa93,wh94,ma02,no16,dr18,sh19}, which showed that the object is variable on short and long timescales. Reverberation mapping analysis showed that the Balmer lines respond with a delay of $\sim$10--15 days to the continuum variations \cite[][]{wa93,dr18,sh19} with even shorter delays in the line wings \citep{dr18}. NGC 3516  was recently confirmed to be an optical CL AGN as a result of a long-term ($\sim$22 years, 1996--2018) optical monitoring campaign \citep{sh19}. The campaign showed that the broad Balmer emission lines almost completely disappeared in 2014, and that at the end of the campaign in 2018 a weak, blueshifted and asymmetric broad component started to reappear \citep{sh19}. This AGN is known to show dramatic optical variability and has changed its type in the past several times \cite[][]{as68,sh19}. However, it has been in a low-activity state in the optical since 2013, with some rebrightening occurring at the end of 2015 until the beginning of 2016, after which it remained in a dormant state. For NGC 3516, it is well known that the broad emission lines show complex, multicomponent profiles, indicating complex kinematics of the BLR, with evidence of outflowing, infalling, and a disk-like emitting region that mostly contributes to the line wings \cite[e.g.,][]{po02,de10,sb17, dr18}. 

NGC 3516 shows even more interesting properties in the high-energy band \cite[e.g.,][]{ed00,ne02,ma02,tu05,me10,li14,hu14,no16}. \citet{ed00} showed that there are uncorrelated trends on longer timescales between the X-ray and optical bands, which was supported with the later findings of \citet{ma02}. This is not unusual for Seyfert galaxies, as discussed by \citet{ed00}. Recently, \citet{no16} found from a  simultaneous X-ray and optical monitoring program during the faintest phase of NGC 3516 (2013--2014) that the X-ray flux and B-band variations were comparable and significantly correlated, with the B-band delayed by $\sim$2 days relative to the X-rays. 
Both results indicate that the standard X-ray reprocessing model in which the X-ray source heats a stratified accretion disk, which then re-emits in the optical and ultraviolet, may not be applicable in the case of NGC 3516 and that these two emitting regions are powered primarily by different processes \citep{ed00,no16}.
%, which suggests that the disk may be truncated and characterized by a radiatively inefficient accretion flow .
Also, NGC 3516 is known for the presence of complex absorption, showing the presence of eight kinematic components in the UV \citep{kr02}, supported by at least three distinct outflowing components in the X-rays \cite[][]{me10,tu11,hu14}. Interestingly, \citet{me10} concluded, similarly to \citet{ne02}, that the X-ray variability is better understood as a consequence of changes in the source continuum emission than in the warm absorber.

Here we report on the very recent flare discovered during the latest intensive photometric monitoring in \textit{U} and \textit{B} filters that may be an indicator that NGC 3516 is transitioning from a low to a high state of activity. We also acquired an optical spectrum during this period, covering H$\beta$ and H$\alpha$ lines, in order to find differences between their profiles (intensities) with respect to the latest published ones \citep{sh19}. In order to analyze the previous episodes of violent variability, our optical observations are supported with the archival UV data of \textit{Swift}/Ultraviolet/Optical Telescope (UVOT) and X-ray \textit{Swift}/XRT data of NGC 3516. This paper is organized as follows: in Section 2, we describe observations and data reduction; in Section 3, we list the main results; in Section 4, we discuss our findings; and in Section 5, we outline our conclusions and future work.

 %  \citet{mizuno} determined
 % of Baker's (\citeyear{baker}) standard one-zone model. 
% (Tscharnuter \citeyear{tscharnuter}, Balluch \citeyear{balluch}),

%                                     Two column figure (place early!)
%______________________________________________ Gamma_1 (lg rho, lg e)
   \begin{figure*}
   \centering
   \includegraphics{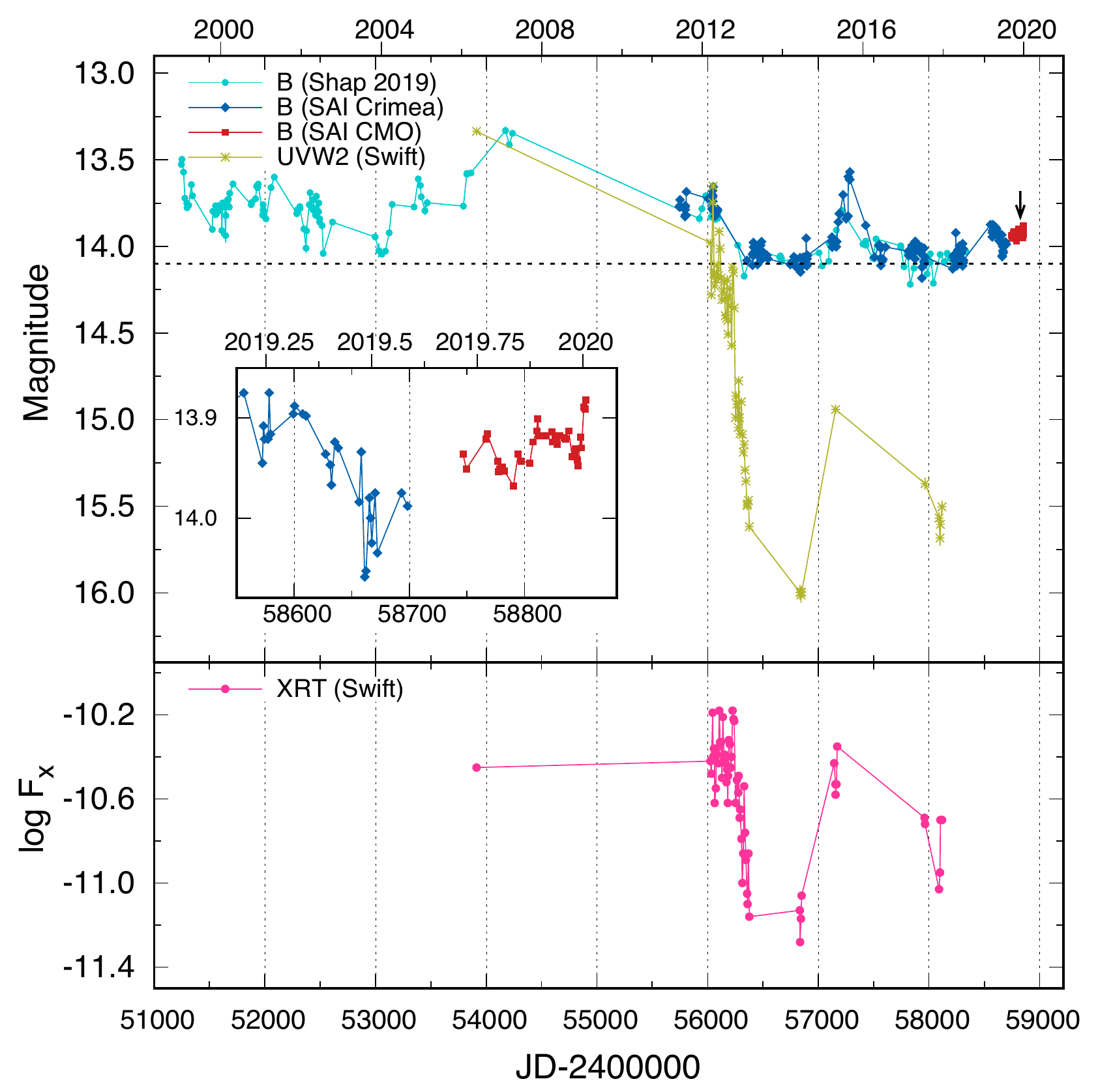}
   \caption{Multiwavelength light curves of NGC 3516. (Top panel) Long-term photometric optical and UV light curves in \textit{B} and \textit{UVW2} filters: circles denote data from \citet{sh19}, squares denote the result from this paper (SAI CMO), diamonds stand for the data from SAI Crimea campaign shown for continuity, and asterisks denote \textit{UVW2} UVOT \textit{Swift} photometry. All magnitudes are reduced to the same aperture of 10\arcsec. A dashed horizontal line at magnitude 14.1 is given at the bottom of the figure to guide the eye, and the arrow indicates epochs when spectra were acquired. (Bottom panel) \textit{Swift}/XRT 0.5-10 keV X-ray flux light curve in erg cm$^{-2}$ s$^{-1}$.
   }
              \label{lc_long}%
    \end{figure*}
%

%__________________________________________________________________

\section{Observations and data reduction}

\subsection{Photometry}

The intensive photometric monitoring was performed in \textit{U} and \textit{B} filters from 2019 Sept 16 to 2020 Jan 04 with the RC600, a new automated telescope of the Caucasus Mountain Observatory at the Sternberg Astronomical Institute (SAI CMO) of Moscow State University. The RC600 has a 600-mm main mirror with a focal length of 4200 mm and is equipped with an Andor iKon-L BV camera (installed filters \textit{U, B, V, Rc, Ic, g', r', i'}). The field of view is 22\arcmin$\times$22\arcmin, with scaling of 0.67\arcsec \, per pixel \cite[for more details see][]{be20}. During these 4 months, data were obtained during good photometric nights with several exposures (2-6) in each band, with typical exposure times of 300s in \textit{U} and 20-30 s in \textit{B} filter. The photometry was done with a small aperture of 6.7\arcsec \, in diameter, and using comparison stars from \citet{ld93}. 

To compare the new data with the long-term trend in NGC 3516, we use: (i) data from the \textit{UBV} photoelectric monitoring initiated in the 1970s \cite[see details in][]{ly72}, with the aperture of 14.3\arcsec \, in diameter using a photoelectrical photometer with EMI 9789 photomultiplier attached to a a 60cm telescope in the Crimean astronomical station SAI MSU \cite[the same as was used by][]{ld93}, and the same comparison stars as in this campaign, and (ii) data from the long-term campaign with the aperture size of 10\arcsec , which are described in \citet{sh19}. To match these two long-term light curves, we used the overlapping points to reduce all data to the same aperture of 10\arcsec. The long-term photometric light curve in \textit{B} filter is presented in Fig. \ref{lc_long}.

%                                                One column figure
%----------------------------------------------------------- S_vib
   \begin{figure}
   \centering
   \includegraphics[width=\hsize]{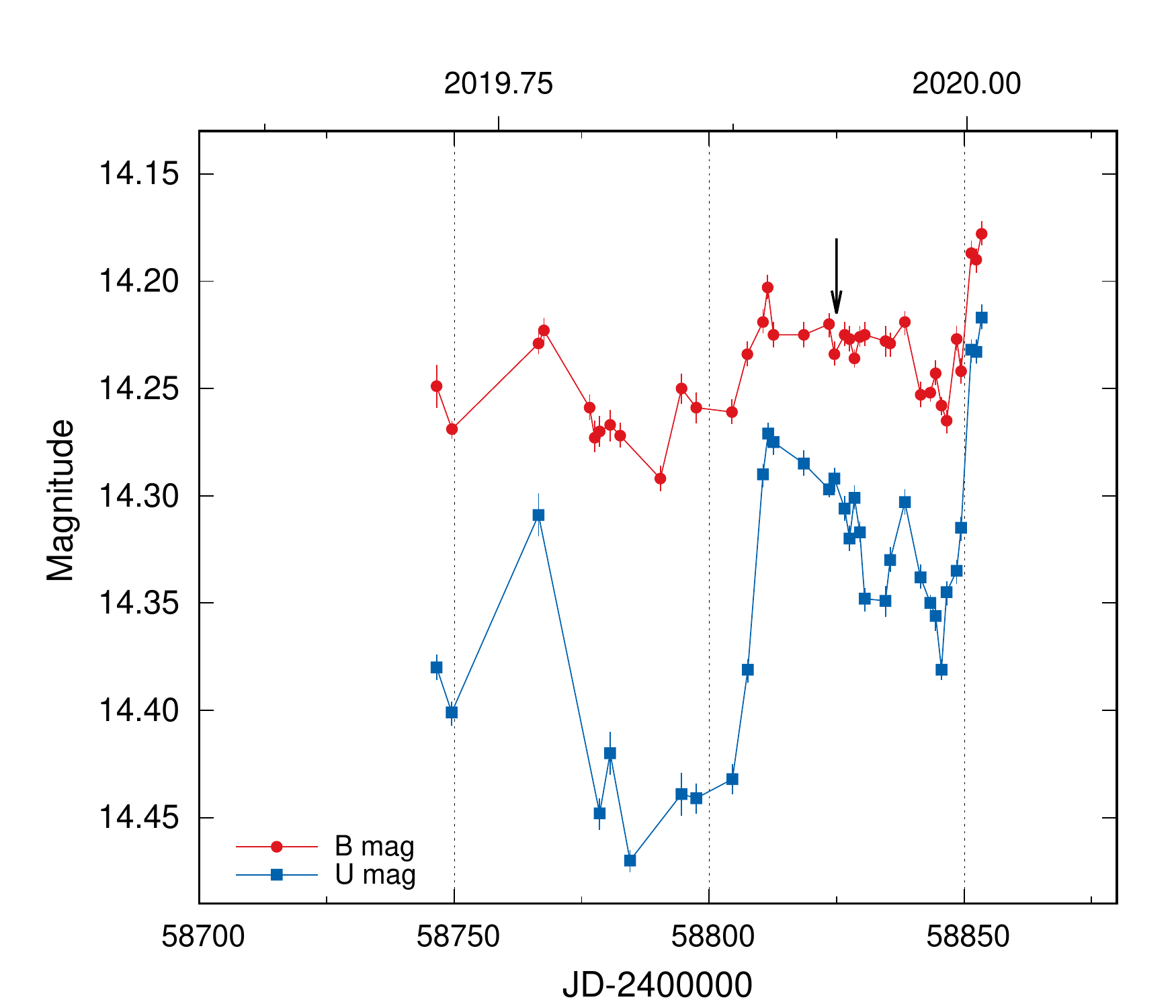}
      \caption{Short-term light curves in \textit{U} and \textit{B} bands (aperture size 6.7\arcsec) from the high-cadence monitoring campaign from 2019 Sept 16 to 2020 Jan 4. The arrow indicates epochs when spectra were acquired.
              }
         \label{lc_UB}
   \end{figure}
%
%______________________________________________________________

\subsection{Spectroscopy}

The new spectra were obtained on 2019 Dec 07 (JD 2458824.57) and 2019 Dec 09 (JD 2458826.59), with the 2.5m telescope of the SAI CMO equipped with the medium-resolution optical double-beam spectrograph, the Andor Newton 940P cameras with CCDs E2V CCD42-10, and the volume phase holographic grating. The slit width was 1\arcsec \, and 1.5\arcsec \,for 2019 Dec 07 and 2019 Dec 09, with exposure times of 600s and 1000s, respectively, giving a spectral resolution of $\sim$4\AA \, and S/N ratio of 30-50 near H$\beta$, and $\sim$3\AA \, and S/N ratio of 40-90 near H$\alpha$. The spectrophotometric data reduction was carried out using our python3 scripts, including bias, flat-field, and dark corrections, cosmic ray removal, two-dimensional wavelength linearization, background subtraction, and relative flux calibration based on spectrophotometric standard stars.

% model of \citet{baker},
% \begin{itemize}
%      \item hydrostatic equilibrium,
%      \item thermal equilibrium,
%      \item energy transport by grey radiation diffusion.
%   \end{itemize}

%(Eqs.\ (34a,\,b,\,c) in Baker \citeyear{baker}). 

%  \begin{enumerate}
%      \item a factor containing local timescales only,
%      \item a factor containing only constitutive .
%   \end{enumerate}

%\emph{  functions of the thermodynamic state in the local zone}. 

%__________________________________________________ One column table
%   \begin{table}
%      \caption[]{Measured photometric magnitudes of NGC 3516. Columns are:(1) date; (2) modified Julian date (MJD); (3) mean seeing in arcsec; (4)–(5) UB magnitudes. The full table is available online as supporting material.}
%         \label{obs_log}
%     $$ 
%         \begin{array}{p{0.2\linewidth}lccc}
%            \hline
%            \noalign{\smallskip}
%            Date      &  {\rm JD}  & {\rm seeing} & U & B \\
%            & {[240+]} & {\rm [arcsec]} & & \\
%            \noalign{\smallskip}
%            \hline
%            \noalign{\smallskip}
%            23 Sep 2019 & 58746.5 & ? & 14.380 & 14.249 \\
%            26 Sep 2019 & 58749.5 & ? & 14.401 & 14.269 \\
%            ... & ... & ...  \\ 
%            04 Jan 2020 & 58853.5 & ? & 14.217 & 14.178  \\
%            \noalign{\smallskip}
%            \hline
%         \end{array}
%     $$ 
%   \end{table}
%
  
%----------------------------------------------------------- S_vib
   \begin{figure}
   \centering
   \includegraphics[width=\hsize]{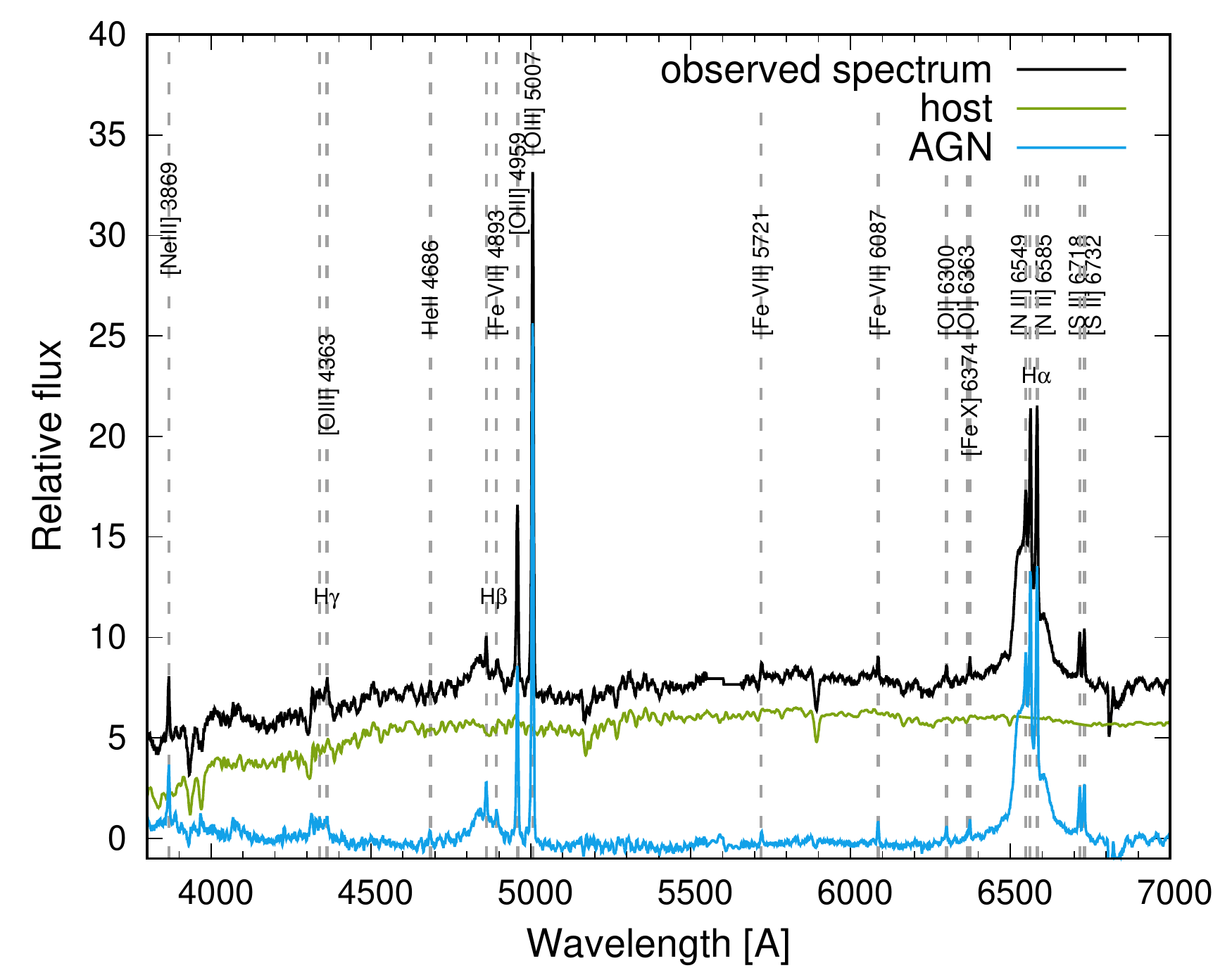}
      \caption{Host galaxy fitting using PCA in the case of the spectrum observed on 2019 Dec 07. The position of all prominent emission lines is marked.
              }
         \label{host}
   \end{figure}
%
%______________________________________________________________

\subsection{Archival UV and X-ray data}

The {\it Neil Gehrels Swift Observatory} \citep{ge04} provided simultaneous XRT and UVOT monitoring of NGC 3516 from 2006 to 2018, with the most intensive campaign taking place in 2012--2014. Some of the {\it Swift} data used in our study were previously published by \cite{bu17}. In order to trace the behavior of NGC 3516 on a longer timescale, data from both the XRT and UVOT telescopes were re-reduced consistently from the raw data available in the archive \cite[for details see][]{bu17,ok17a}. The X-ray spectra were obtained through the standard online tools provided by the UK Swift Science Data Centre\footnote{\url{http://www.swift.ac.uk/user_objects/}} \citep[][]{ev09}. To account for low count statistics, the XRT spectra in the 0.5-10 keV range were binned to assure at least one count per energy bin and fitted using the W-statistic \cite[][]{wa79}. To get the source flux in physical units, spectra were fitted with a simple absorbed power-law model leaving the photon index free to vary and freezing the equivalent absorption column $N_{\rm H}$ at $3\times10^{20}$ cm$^{-2}$, the Galactic column density along the line of sight to the galaxy \citep{ka05}.

The image analysis of the {\it Swift}/UVOT data in different bands (\textit{V, B, U, UVW1, UVW2, UVM2}) was done again following the procedure provided by the UK Swift Science Data Centre. Photometry was performed with the {\sc uvotsource} tool with aperture diameter of 10\arcsec\ and 40\arcsec\ for the source and background, respectively. The archive provides the largest number of epochs for the \textit{UVW2} magnitude. The background was chosen with the center about 2\arcmin\ away from the galaxy for all filters. The resulting light curves of the XRT X-ray flux in the 0.5-10 keV band and the UVOT \textit{UVW2} magnitude are shown in Fig. \ref{lc_long}.

%----------------------------------------------------------- S_vib
   \begin{figure}
   \centering
   \includegraphics[width=\hsize]{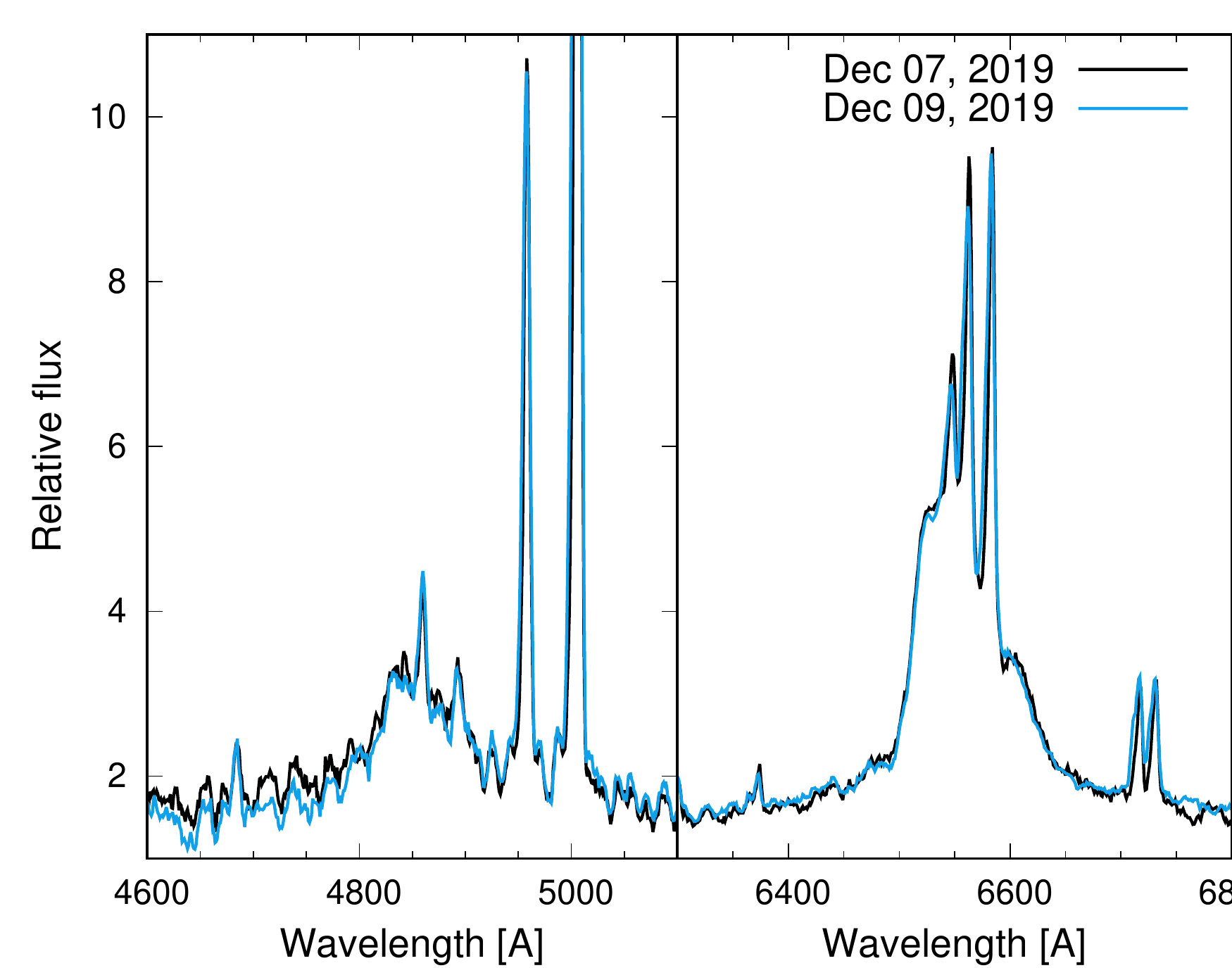}
      \caption{Two spectra observed in 2019, the H$\beta$ (left) and H$\alpha$ (right) spectral regions are shown. We note that the spectra are with the original different spectral resolution and are scaled for comparison.
              }
         \label{Hab}
   \end{figure}
%
%______________________________________________________________

\section{Analysis and results}

From the long-term light curve in Fig. \ref{lc_long}, it is clear that NGC 3516 experienced several low and high-activity states. The object was the brightest in 2007, followed with the minimum in 2014 \citep{sh19}. Several weaker changes in the optical flux happened afterwards, with the strongest subsequent flare occurring at the end of 2015 when the broad component of H$\alpha$ line started to increase \cite[see Figure 12 in][]{sh19}, after which the object entered the low-activity state. The latest brightening probably started from the end of 2018, followed by the local minimum at the end of July 2019, after which there was again brightening with the maximum in January 2020. This is more obvious when we plot only the latest data, now both in \textit{U} and \textit{B} filters (Fig. \ref{lc_UB}). The accuracy of photometric measurements extracted from the small-size aperture allows for the short-term variability to be seen. The increase in the \textit{U} magnitude is more prominent, and is 0.25 mag, whereas in the \textit{B} filter, it is 0.11 mag (Fig. \ref{lc_UB}). Figure 2 shows the trend of increase, and compared to the minimum seen in the first half of 2019 (Fig. \ref{lc_long}) the change in the \textit{B} magnitude (scaled to 10\arcsec \, aperture) is $\sim$0.15. The variability amplitude appears lower in the larger-size aperture, therefore the gradient in the smaller-size aperture is probably larger.

The \textit{U} and \textit{B} magnitudes are closely correlated. However, the strong increase in the \textit{U} magnitude, which could be interpreted as the continuum emission mostly coming from the accretion disk with the contribution of the Balmer continuum coming from the BLR, does not have the same amplitude in the \textit{B} magnitude, which  also contains the H$\beta$ line. This is most likely due to the host-galaxy contribution, which is much stronger in the \textit{B} band. 
Looking at the light curves it seems there is little or no time delay between the \textit{U} and \textit{B} band. A more detailed analysis of the light-curve properties \cite[e.g., time delay and oscillations; see][]{ko18} will be given in a separate publication.

%We obtained preliminary estimate of the time-lag of $\sim$1 day between U and B bands, using the GPccf method, which is utilizing generalized Gaussian processes to model the observed light curves and the zDCF to extract the time-delays \citep[see description in][]{ko15,ko18,sh19}. But a more detailed analysis will be given elsewhere.

The most striking change in the flux is seen when the UV and X-ray data are considered (Fig. \ref{lc_long}). In the 2012-2014 intensive monitoring campaign, the \textit{UVW2} magnitude changed by more than 2 mag, while \textit{B} changed by only $\sim$0.3 mag. The largest amplitude in the available data was $\sim$3 mag from 2006 to 2015. The X-ray data show similar behavior; that is, the difference between the minimum in 2014 and a rise in 2016 in the optical band is very small, on the order of $\sim$0.2 mag, while the X-ray flux changed significantly by about a factor of five. Most importantly, the optical variations are similar and  closely follow the trends seen in the UV and X-ray, therefore we may expect that in the UV/X-ray bands the object is experiencing a much more abrupt transition.

We obtained the optical spectra of NGC 3516 in order to follow the flare spotted in the photometric data. To extract the broad line profile, we subtracted the host-galaxy spectrum using the principal component analysis (PCA), a statistical method that was described in \cite{vb06}, and was used in the previous campaign \cite[for details see][]{sh19}. Figure \ref{host} shows the result of the host galaxy fitting for the spectrum obtained on Dec 09, 2019. The positions of prominent lines are also marked in Fig. \ref{host}.

After subtracting the host galaxy, several forbidden high-ionization lines are clearly seen. The strongest are [\ion{Fe}{vii}] $\lambda$4893, $\lambda$5721, $\lambda$6087, [\ion{Fe}{x}] $\lambda$6374, and [\ion{Ne}{ii}] $\lambda$3869.  These so-called coronal lines are probably originating from the inner narrow-line region or from the inner edge of the torus \cite[e.g.,][]{pe88,ro15}, and for their creation most likely a hot X-ray wind is needed \citep{pv95} or some special geometry, like polar conical regions \citep{ok91}. The variability of coronal lines has been detected before \cite[e.g.,][]{la15a,la15b}, and their strengthening has been shown to be a signature of transition from a low to a high-activity state in AGNs \citep{ok19b}. Several other examples of CL AGNs that have displayed variable coronal line emission are known, such as for example J000904.54-103428.7, J013203 \citep{ml19}, NGC 5548 \citep{la15b,fa16}, Mkn 110 \citep{kb02}, NGC 4151 \citep{ok91,la15a}, and ZTF18aajupnt \citep{fr19}.

%----------------------------------------------------------- S_vib
   \begin{figure}
   \centering
   \includegraphics[width=9.5cm]{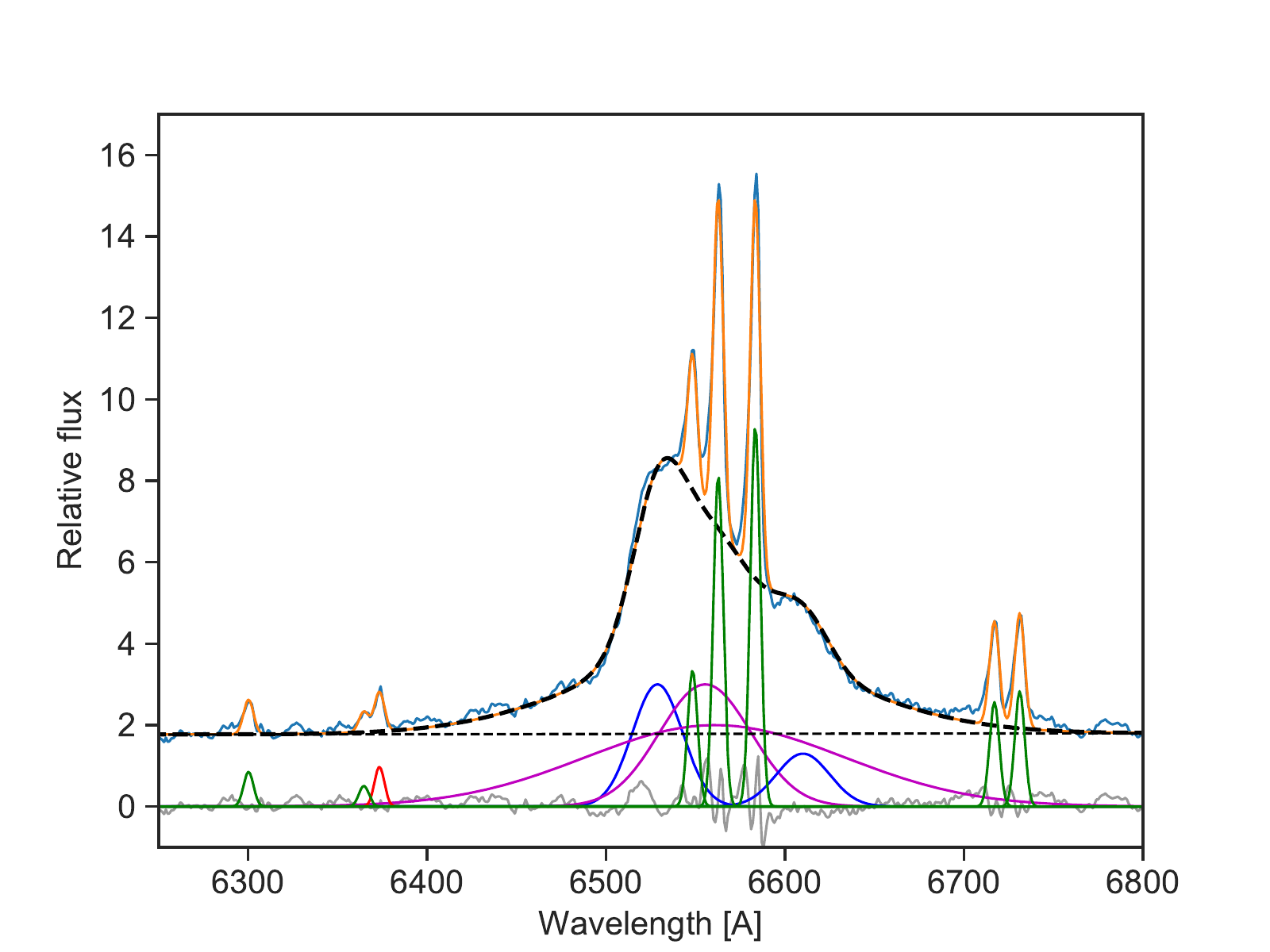}
   \includegraphics[width=9.5cm]{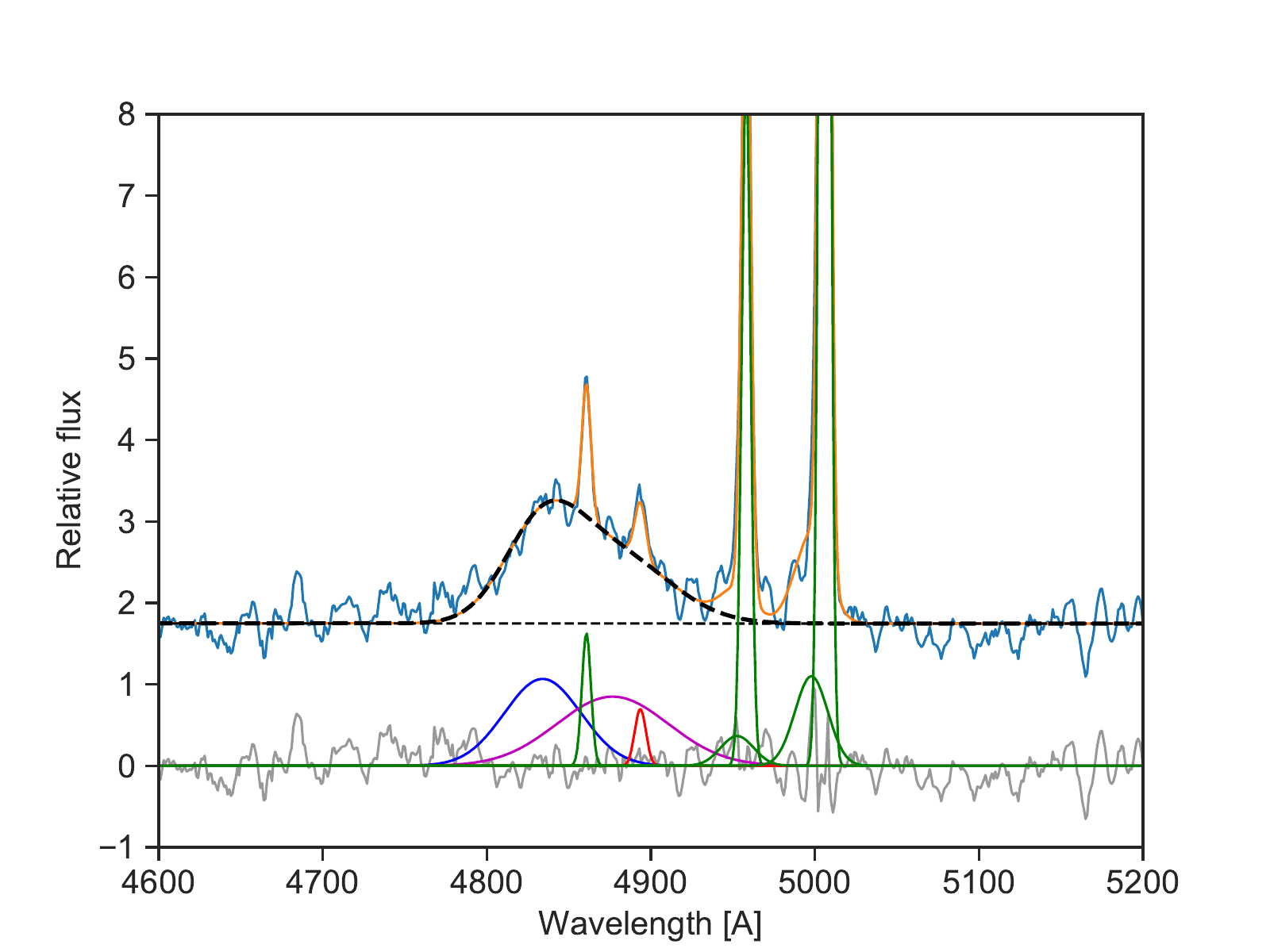}
      \caption{Multicomponent fitting of the H$\alpha$ (up) and
      H$\beta$ (bottom) lines. Below the observed (blue) and modeled (orange) spectrum, the residual (gray) and all Gaussian components are shown (blue, magenta - broad components of H$\alpha$ (up) and
      H$\beta$; green - narrow lines; red - forbidden [\ion{Fe}{x}] $\lambda$6374 and [\ion{Fe}{vii}] $\lambda$4893 lines). The thick dashed line indicates the modeled broad line; this is the sum of all broad Gaussian components. 
              }
         \label{gauss}
   \end{figure}
%
%______________________________________________________________

Furthermore, we modeled the broad line profiles with multiGaussian components \cite[following the procedures and parameter criteria described in e.g.,][and references therein]{po04,il06,di07,ko10,sh12,po14}, using the newly developed python-based AGN line-fitting code \cite[see][]{rakic}. This code simultaneously fits the underlying continuum and all emission lines, and is based on the Sherpa python package \citep{db19}. In Fig. \ref{gauss} we plot the result of the multicomponent fitting of the H$\alpha$ (up) and H$\beta$ (bottom). Due to its complex profile, the H$\alpha$ broad line is modeled with four components (blue and magenta lines in Fig. \ref{gauss}), whereas H$\beta$ was reproduced with only two components due to its low S/N. Both lines show asymmetric profiles with a blueshifted peak and very extended wings that are better seen in H$\alpha$ line because of the higher S/N ratio. 

The blended profile of [O I] $\lambda$6363 and [\ion{Fe}{x}] $\lambda$6374 was also fitted (see Fig.\ref{gauss}, upper panel). From the fit, the estimated ratio of [O I]+[\ion{Fe}{x}] to [O I] $\lambda$6300 is $\sim$1.2, similar that observed in NGC 1566 during the brightening phase \citep{ok19b}, but we note that it is strongly dependent on the estimated level of the underlying continuum and H$\alpha$ extended wing. From the clear single profile of [\ion{Fe}{vii}] $\lambda$6087 (see Fig.\ref{host}), we estimated that the line fluxes have increased by the factor of approximately two compared to the last published data of \citet{sh19}. The presence of strong coronal [\ion{Fe}{vii}] and [\ion{Fe}{x}] lines could be connected with the TDEs \cite[e.g.,][]{ya13} but more intense [\ion{Fe}{x}] would be expected \cite[][]{wa12}. On the other hand, their strengthening can be a tracer of an awaking phase, as recently shown in a "turn-on" of broad-line AGNs detected in a normal galaxy \citep{ya19}.

%----------------------------------------------------------- S_vib
   \begin{figure}
   \centering
   \includegraphics[width=\hsize]{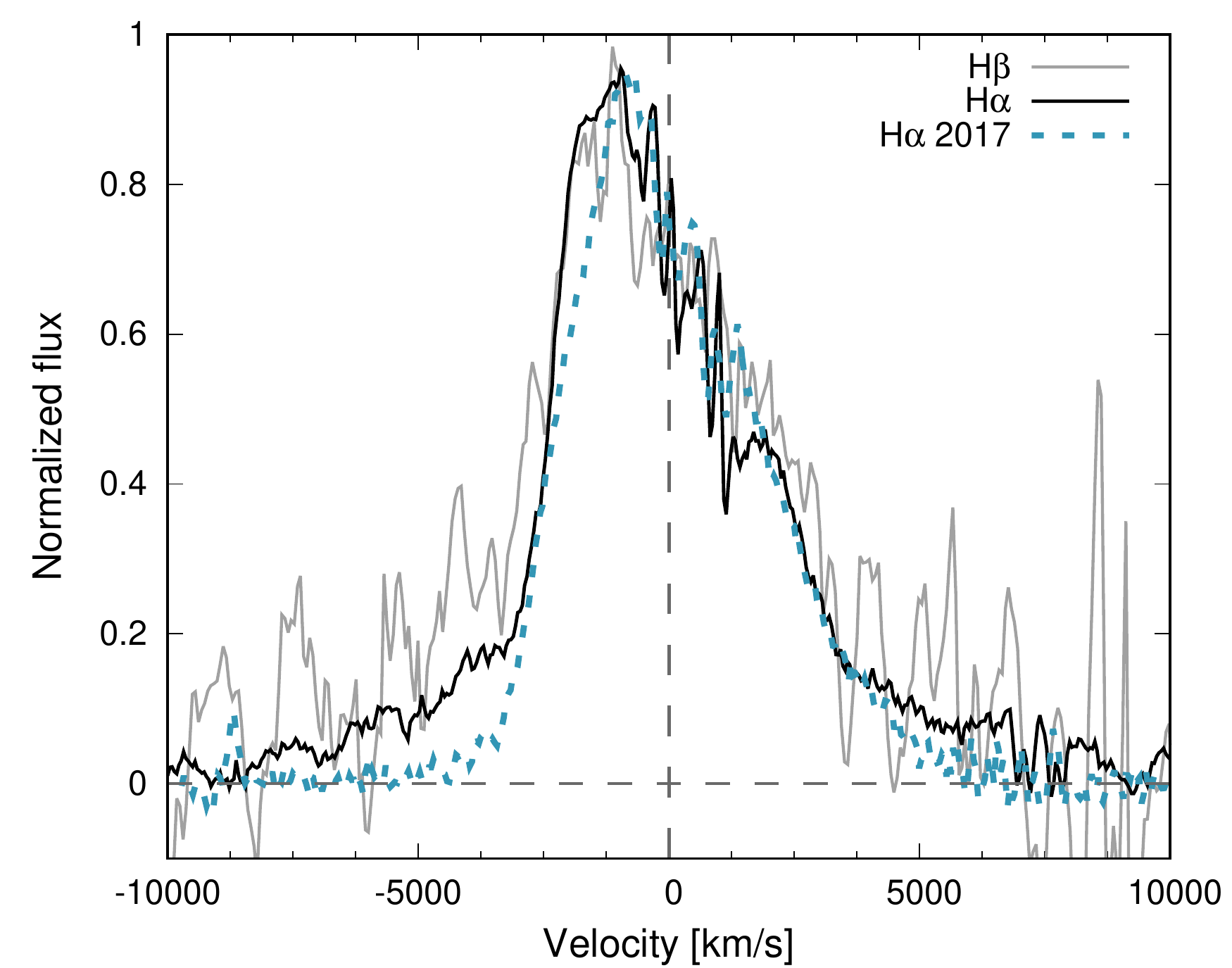}
      \caption{Comparison of the normalized H$\alpha$ and
      H$\beta$ broad-line profiles from 2019. The wavelength has been converted to the velocity scale. The best-quality broad-line profile of H$\alpha$ from the \cite{sh19} campaign obtained in February 2017 is also plotted.
              }
         \label{broad}
   \end{figure}
%
%______________________________________________________________

%Pier & Voit 1995)

%Why this is given here, and could it be compared with other papers?

Figure \ref{broad} shows the broad-line profiles of H$\alpha$ and H$\beta$ lines; these have been normalized for comparison. The lines show the same double-peaked profile that is better seen in the H$\alpha$ line because of higher S/N in the spectra. They have a similar full width half maximum (FWHM) of $\sim 3400$ km s$^{-1}$ (Fig. \ref{broad}). Compared to the results from our previous long-term campaign (broad H$\alpha$ profile is also shown in Fig. \ref{broad}), the FWHM is on the order of the average FWHM during the whole 22-year campaign \citep{sh19}. Both broad components are still significantly shifted to the blue, as was noticed in 2017 when the broad component started to appear. The lines were then blueshifted for around 1000 km s$^{-1}$, and it may be that the blue peak has changed its position, which requires verification with newer continuous spectral observations. In the H$\alpha$ profile, signatures of the blue and red bumps are noticeable, whereas in both lines the very broad component of  $\sim$10,000 km s$^{-1}$ in width seems to appear. These features indicate that the BLR of NGC 3516  is complex and that broad line emission  is present both from the outflowing material and the disk, as suggested by \cite{po02}.

\section{Discussion}

During the high-cadence fourth-month optical monitoring campaign of the prototypical changing-look Seyfert galaxy NGC 3516, we detected an increase in the optical photometric light curves and spectral features including the brightening of coronal Fe lines and a proposed emergent broad Balmer feature. We strengthened our analysis with previous optical photometric data (1999--2018), as well as with \textit{Swift} X-ray and UV archival monitoring data of NGC 3516 from 2006--2018.

We find that the change in the \textit{U} and \textit{B} band is $\sim$ 0.25 mag and 0.11 mag, respectively. This could be considered as ordinary broad-line AGN variability \cite[e.g.,][]{vb04,se07} and one could expect  an increase of around $\sim$1 mag or even more \cite[e.g.,][]{gr17,gr20,ml19,ru18} in the case
of a peculiar flaring (or an extreme brightening). However, a high increase of $\sim$1 mag is more likely to happen in bright quasars, and this more strict limit is used as a flaring criterion in large surveys, such as for example the Sloan Digital Sky Survey in which the magnitude uncertainties can be about 0.2 mag or more. However, here we performed high-precision photometry toward NGC 3516 achieving $\sim$0.01 mag uncertainties, and therefore we are able to detect real magnification in the object brightness. Moreover, one would expect the changes in the \textit{U} and \textit{B} of the CL AGN to be significantly smaller than that in high-energy bands, i.e. the dramatic change in the X-ray and far UV band is probably followed by a much lower change in the optical, since the host galaxy has a very strong contribution in the optical spectral band. This is clearly seen in the case of NGC 1566, which is the closest CL AGN \citep{ok19b}. At the distance of NGC 3516, using the same aperture as for NGC 1566, there is a significantly higher contribution of the host galaxy to the NGC 3516 optical flux. Comparing the NGC 1566 dramatic variations of $\sim$1 mag observed in 2018 \citep[see][]{ok19b} with the NGC 3516 ones reported in this paper, if NGC 1566 were at the same distance as NGC 3516, then for the same aperture the NGC 1566 \textit{B}-band brightness would change by much less than 0.2 mag. Furthermore, we would not be able to detect these large variations of $\sim$1 mag, but only a smaller change of 0.2--0.3 mag. This suggests that real  small-amplitude optical variations in NGC 3516 could be supported with much stronger variations of a few magnitudes in the X-ray and far UV band, and could indicate that the object is going through a transition. Therefore, further multiwavelength observations are required to distinguish that this optical variability is beyond ordinary AGN behavior and that it could be linked to a change in the state of AGN activity.

One of the signatures of a transition from  low to high activity is the presence of strong coronal lines, of which [\ion{Fe}{x}] is of particular interest. Intense X-ray flux is needed for this coronal line to be strong, because we need 0.2 keV energy photons to ionize Fe$^{8+}$ to Fe$^{9+}$. We detected this line earlier,  in our long-term spectroscopic campaign \cite[see, e.g. Figures 2 and 3 in][]{sh19}, although it was not mentioned in the publication. It appears that in the highest state of activity in 2007 \cite[see Figure 12 in][]{sh19} the line was not particularly prominent. This could be due to  either the poorer quality of the spectra, and the fact that H$\alpha$ is much stronger, making it difficult to detect [\ion{Fe}{x}], or some physical phenomenon, which future monitoring of coronal-line variability will reveal. 

%In the very high level of X-ray the ionization condition may change significantly, so this could be the reason why [\ion{Fe}{x}] is \textbf{weaker}, and moreover the H$\alpha$ is is much stronger and makes it difficult to detect [\ion{Fe}{x}].

%It is possible to find publications about that for references.

\section{Conclusions}

The changing-look AGN NGC 3516 was monitored intensively in \textit{U} and \textit{B} filters over a period of four months, and these observations were supported with additional spectral observations once the increase in brightness was spotted. We performed accurate photometric measurements and performed a preliminary analysis of the line profiles. We support our analysis of the past behavior of NGC 3516 with archival \textit{Swift} data in UV and X-ray bands. Our main findings can be summarized as follows.

  \begin{enumerate}[(i)]
      \item We detected a general increase in brightness compared to the last three years of photometric data. The latest flare, at the end of 2019, is seen in both \textit{U} and \textit{B} magnitude to a maximum amplitude of 0.25 mag and 0.11 mag, respectively. The variations in the \textit{U} and \textit{B} magnitude are well correlated, but the amplitude of variability in \textit{U} is significantly greater than in \textit{B,} which is partially connected with the stronger contribution of the host galaxy in the \textit{B} band.
      \item Several coronal lines, such as the forbidden high-ionization iron lines [\ion{Fe}{vii}] and [\ion{Fe}{x}], are clearly detected in the NGC 3516 spectrum observed in 2019, and increased in flux compared to the end of \citet{sh19} campaign. The strengthening of these lines may indicate that the AGN is entering a more active phase \cite[see, e.g.,][]{ya19}.
      \item The broad Balmer lines show a complex double-peaked structure, with an asymmetric profile and prominent blue peak \cite[similar to in][]{sh19}. However, it seems that a very broad component of $\sim$10,000 km s$^{-1}$ width is starting to appear, which may also indicate that the AGN is experiencing a transition to a higher activity phase.
      \item The optical variations are similar and  closely follow the trends seen in UV and X-rays, and therefore we may expect that in UV and X-rays the object is experiencing a more abrupt transition. However, it is not sufficiently empirically demonstrated that a flare in the X-rays is coupled to a transition in the presence or absence of emission lines \citep{la17}, and therefore a final conclusion may only be possible with future observations and additional analysis.
   \end{enumerate}

The photometric and spectroscopic observations presented in this paper indicate that the AGN of NGC 3516 may be in a transition phase, changing from a low-activity state to one of high activity. Further intensive multiwavelength (optical, UV, and X-ray) monitoring in photometry and spectroscopy is needed and may help to elucidate the processes behind the changing-look phenomenon in AGNs.  

%For sure, NGC 3516 continues to be one of the intriguing AGN and will be further continuously monitored in photometry and spectroscopy by our campaign.

\begin{acknowledgements}
      This work made use of data supplied by the UK Swift Science Data Centre at the University of Leicester. Part of this work was supported by the Ministry of Education, Science and Technological Development of the Republic of Serbia through the project number 176001, and from the M.V. Lomonosov Moscow State University Program of Development (RC600 \& TDS). BAA, EOM, SAP, NIS, AVD acknowledge the support by the Program of development of M.V. Lomonosov Moscow State University (Leading Scientific School 'Physics of stars, relativistic objects and galaxies'). EOM, NIS, AVD, AMT, SGJ are also supported by the RSF grant 17-12-01241. ST acknowledges financial support from the Russian Foundation for Basic Research project 17-52-80139 BRICS-a. We also express our thanks to the anonymous referee for helpful comments and suggestions that greatly improved this work.
\end{acknowledgements}

\bibliographystyle{aa} % style aa.bst
\bibliography{37532corr_REV} % your references Yourfile.bib

\begin{thebibliography}{71}
\expandafter\ifx\csname natexlab\endcsname\relax\def\natexlab#1{#1}\fi

\bibitem[{{Andrillat} \& {Souffrin}(1968)}]{as68}
{Andrillat}, Y. \& {Souffrin}, S. 1968, \aplett, 1, 111

\bibitem[{{Berdnikov} {et~al.}(2020){Berdnikov}, {Belinski}, {Shatsky},
  {Burlak}, {Ikonnikova}, {Mishin}, {Cheryasov}, \& {Zhuyko}}]{be20}
{Berdnikov}, L.~N., {Belinski}, A.~A., {Shatsky}, N.~I., {et~al.} 2020,
  Astronomy Reports, 64, 310

\bibitem[{{Buisson} {et~al.}(2017){Buisson}, {Lohfink}, {Alston}, \&
  {Fabian}}]{bu17}
{Buisson}, D.~J.~K., {Lohfink}, A.~M., {Alston}, W.~N., \& {Fabian}, A.~C.
  2017, \mnras, 464, 3194

\bibitem[{Burke {et~al.}(2019)Burke, Laurino, dtnguyen2, Budynkiewicz,
  Aldcroft, Siemiginowska, Deil, wmclaugh, Sipocz, \& Leinweber}]{db19}
Burke, D., Laurino, O., dtnguyen2, {et~al.} 2019, sherpa/sherpa: Sherpa 4.11.1

\bibitem[{{Campana} {et~al.}(2015){Campana}, {Mainetti}, {Colpi}, {Lodato},
  {D'Avanzo}, {Evans}, \& {Moretti}}]{ca15}
{Campana}, S., {Mainetti}, D., {Colpi}, M., {et~al.} 2015, \aap, 581, A17

\bibitem[{{De Rosa} {et~al.}(2018){De Rosa}, {Fausnaugh}, {Grier}, {Peterson},
  {Denney}, {Horne}, {Bentz}, {Ciroi}, {Dalla Bont{\`a}}, {Joner}, {Kaspi},
  {Kochanek}, {Pogge}, {Sergeev}, {Vestergaard}, {Adams}, {Antognini}, {Araya
  Salvo}, {Armstrong}, {Bae}, {Barth}, {Beatty}, {Bhattacharjee}, {Borman},
  {Boroson}, {Bottorff}, {Brown}, {Brown}, {Brotherton}, {Coker}, {Clanton},
  {Cracco}, {Crawford}, {Croxall}, {Eftekharzadeh}, {Eracleous}, {Fiorenza},
  {Frassati}, {Hawkins}, {Henderson}, {Holoien}, {Hutchison}, {Kellar},
  {Kilerci-Eser}, {Kim}, {King}, {La Mura}, {Laney}, {Li}, {Lochhaas}, {Ma},
  {MacInnis}, {Manne-Nicholas}, {Mason}, {McGraw}, {Mogren}, {Montouri},
  {Moody}, {Mosquera}, {Mudd}, {Musso}, {Nazarov}, {Nguyen}, {Ochner},
  {Okhmat}, {Onken}, {Ou-Yang}, {Pancoast}, {Pei}, {Penny}, {Poleski},
  {Portaluri}, {Prieto}, {Price-Whelan}, {Pulatova}, {Rafter}, {Roettenbacher},
  {Romero-Colmenero}, {Runnoe}, {Schimoia}, {Shappee}, {Sherf}, {Simonian},
  {Siviero}, {Skowron}, {Skowron}, {Somers}, {Spencer}, {Starkey}, {Stevens},
  {Stoll}, {Tamajo}, {Tayar}, {van Saders}, {Valenti}, {Villanueva},
  {Villforth}, {Weiss}, {Winkler}, {Zastrow}, {Zhu}, \& {Zu}}]{dr18}
{De Rosa}, G., {Fausnaugh}, M.~M., {Grier}, C.~J., {et~al.} 2018, \apj, 866,
  133

\bibitem[{{Denney} {et~al.}(2014){Denney}, {De Rosa}, {Croxall}, {Gupta},
  {Bentz}, {Fausnaugh}, {Grier}, {Martini}, {Mathur}, {Peterson}, {Pogge}, \&
  {Shappee}}]{de14}
{Denney}, K.~D., {De Rosa}, G., {Croxall}, K., {et~al.} 2014, \apj, 796, 134

\bibitem[{{Denney} {et~al.}(2010){Denney}, {Peterson}, {Pogge}, {Adair},
  {Atlee}, {Au-Yong}, {Bentz}, {Bird}, {Brokofsky}, {Chisholm}, {Comins},
  {Dietrich}, {Doroshenko}, {Eastman}, {Efimov}, {Ewald}, {Ferbey}, {Gaskell},
  {Hedrick}, {Jackson}, {Klimanov}, {Klimek}, {Kruse}, {Lad{\'e}route}, {Lamb},
  {Leighly}, {Minezaki}, {Nazarov}, {Onken}, {Petersen}, {Peterson},
  {Poindexter}, {Sakata}, {Schlesinger}, {Sergeev}, {Skolski}, {Stieglitz},
  {Tobin}, {Unterborn}, {Vestergaard}, {Watkins}, {Watson}, \& {Yoshii}}]{de10}
{Denney}, K.~D., {Peterson}, B.~M., {Pogge}, R.~W., {et~al.} 2010, \apj, 721,
  715

\bibitem[{{Dimitrijevi{\'c}} {et~al.}(2007){Dimitrijevi{\'c}}, {Popovi{\'c}},
  {Kova{\v{c}}evi{\'c}}, {Da{\v{c}}i{\'c}}, \& {Ili{\'c}}}]{di07}
{Dimitrijevi{\'c}}, M.~S., {Popovi{\'c}}, L.~{\v{C}}., {Kova{\v{c}}evi{\'c}},
  J., {Da{\v{c}}i{\'c}}, M., \& {Ili{\'c}}, D. 2007, \mnras, 374, 1181

\bibitem[{{Edelson} {et~al.}(2000){Edelson}, {Koratkar}, {Nandra}, {Goad},
  {Peterson}, {Collier}, {Krolik}, {Malkan}, {Maoz}, {O'Brien}, {Shull},
  {Vaughan}, \& {Warwick}}]{ed00}
{Edelson}, R., {Koratkar}, A., {Nandra}, K., {et~al.} 2000, \apj, 534, 180

\bibitem[{{Elitzur}(2012)}]{el12}
{Elitzur}, M. 2012, \apjl, 747, L33

\bibitem[{{Evans} {et~al.}(2009){Evans}, {Beardmore}, {Page}, {Osborne},
  {O'Brien}, {Willingale}, {Starling}, {Burrows}, {Godet}, {Vetere}, {Racusin},
  {Goad}, {Wiersema}, {Angelini}, {Capalbi}, {Chincarini}, {Gehrels}, {Kennea},
  {Margutti}, {Morris}, {Mountford}, {Pagani}, {Perri}, {Romano}, \&
  {Tanvir}}]{ev09}
{Evans}, P.~A., {Beardmore}, A.~P., {Page}, K.~L., {et~al.} 2009, \mnras, 397,
  1177

\bibitem[{{Fausnaugh} {et~al.}(2016){Fausnaugh}, {Denney}, {Barth}, {Bentz},
  {Bottorff}, {Carini}, {Croxall}, {De Rosa}, {Goad}, {Horne}, {Joner},
  {Kaspi}, {Kim}, {Klimanov}, {Kochanek}, {Leonard}, {Netzer}, {Peterson},
  {Schn{\"u}lle}, {Sergeev}, {Vestergaard}, {Zheng}, {Zu}, {Anderson},
  {Ar{\'e}valo}, {Bazhaw}, {Borman}, {Boroson}, {Brandt}, {Breeveld}, {Brewer},
  {Cackett}, {Crenshaw}, {Dalla Bont{\`a}}, {De Lorenzo-C{\'a}ceres},
  {Dietrich}, {Edelson}, {Efimova}, {Ely}, {Evans}, {Filippenko}, {Flatland},
  {Gehrels}, {Geier}, {Gelbord}, {Gonzalez}, {Gorjian}, {Grier}, {Grupe},
  {Hall}, {Hicks}, {Horenstein}, {Hutchison}, {Im}, {Jensen}, {Jones},
  {Kaastra}, {Kelly}, {Kennea}, {Kim}, {Korista}, {Kriss}, {Lee}, {Lira},
  {MacInnis}, {Manne-Nicholas}, {Mathur}, {McHardy}, {Montouri}, {Musso},
  {Nazarov}, {Norris}, {Nousek}, {Okhmat}, {Pancoast}, {Papadakis}, {Parks},
  {Pei}, {Pogge}, {Pott}, {Rafter}, {Rix}, {Saylor}, {Schimoia}, {Siegel},
  {Spencer}, {Starkey}, {Sung}, {Teems}, {Treu}, {Turner}, {Uttley},
  {Villforth}, {Weiss}, {Woo}, {Yan}, \& {Young}}]{fa16}
{Fausnaugh}, M.~M., {Denney}, K.~D., {Barth}, A.~J., {et~al.} 2016, \apj, 821,
  56

\bibitem[{{Frederick} {et~al.}(2019){Frederick}, {Gezari}, {Graham}, {Cenko},
  {van Velzen}, {Stern}, {Blagorodnova}, {Kulkarni}, {Yan}, {De}, {Fremling},
  {Hung}, {Kara}, {Shupe}, {Ward}, {Bellm}, {Dekany}, {Duev}, {Feindt},
  {Giomi}, {Kupfer}, {Laher}, {Masci}, {Miller}, {Neill}, {Ngeow}, {Patterson},
  {Porter}, {Rusholme}, {Sollerman}, \& {Walters}}]{fr19}
{Frederick}, S., {Gezari}, S., {Graham}, M.~J., {et~al.} 2019, \apj, 883, 31

\bibitem[{{Gehrels} {et~al.}(2004){Gehrels}, {Chincarini}, {Giommi}, {Mason},
  {Nousek}, {Wells}, {White}, {Barthelmy}, {Burrows}, {Cominsky}, {Hurley},
  {Marshall}, {M{\'e}sz{\'a}ros}, {Roming}, {Angelini}, {Barbier}, {Belloni},
  {Campana}, {Caraveo}, {Chester}, {Citterio}, {Cline}, {Cropper}, {Cummings},
  {Dean}, {Feigelson}, {Fenimore}, {Frail}, {Fruchter}, {Garmire}, {Gendreau},
  {Ghisellini}, {Greiner}, {Hill}, {Hunsberger}, {Krimm}, {Kulkarni}, {Kumar},
  {Lebrun}, {Lloyd-Ronning}, {Markwardt}, {Mattson}, {Mushotzky}, {Norris},
  {Osborne}, {Paczynski}, {Palmer}, {Park}, {Parsons}, {Paul}, {Rees},
  {Reynolds}, {Rhoads}, {Sasseen}, {Schaefer}, {Short}, {Smale}, {Smith},
  {Stella}, {Tagliaferri}, {Takahashi}, {Tashiro}, {Townsley}, {Tueller},
  {Turner}, {Vietri}, {Voges}, {Ward}, {Willingale}, {Zerbi}, \&
  {Zhang}}]{ge04}
{Gehrels}, N., {Chincarini}, G., {Giommi}, P., {et~al.} 2004, \apj, 611, 1005

\bibitem[{{Graham} {et~al.}(2017){Graham}, {Djorgovski}, {Drake}, {Stern},
  {Mahabal}, {Glikman}, {Larson}, \& {Christensen}}]{gr17}
{Graham}, M.~J., {Djorgovski}, S.~G., {Drake}, A.~J., {et~al.} 2017, \mnras,
  470, 4112

\bibitem[{{Graham} {et~al.}(2020){Graham}, {Ross}, {Stern}, {Drake},
  {McKernan}, {Ford}, {Djorgovski}, {Mahabal}, {Glikman}, {Larson}, \&
  {Christensen}}]{gr20}
{Graham}, M.~J., {Ross}, N.~P., {Stern}, D., {et~al.} 2020, \mnras, 491, 4925

\bibitem[{{Huerta} {et~al.}(2014){Huerta}, {Krongold}, {Nicastro}, {Mathur},
  {Longinotti}, \& {Jimenez-Bailon}}]{hu14}
{Huerta}, E.~M., {Krongold}, Y., {Nicastro}, F., {et~al.} 2014, \apj, 793, 61

\bibitem[{{Ili{\'c}} {et~al.}(2006){Ili{\'c}}, {Popovi{\'c}}, {Bon},
  {Mediavilla}, \& {Chavushyan}}]{il06}
{Ili{\'c}}, D., {Popovi{\'c}}, L.~{\v{C}}., {Bon}, E., {Mediavilla}, E.~G., \&
  {Chavushyan}, V.~H. 2006, \mnras, 371, 1610

\bibitem[{{Kalberla} {et~al.}(2005){Kalberla}, {Burton}, {Hartmann}, {Arnal},
  {Bajaja}, {Morras}, \& {P{\"o}ppel}}]{ka05}
{Kalberla}, P.~M.~W., {Burton}, W.~B., {Hartmann}, D., {et~al.} 2005, \aap,
  440, 775

\bibitem[{{Kollatschny} \& {Bischoff}(2002)}]{kb02}
{Kollatschny}, W. \& {Bischoff}, K. 2002, \aap, 386, L19

\bibitem[{{Kollatschny} \& {Fricke}(1985)}]{kf85}
{Kollatschny}, W. \& {Fricke}, K.~J. 1985, \aap, 146, L11

\bibitem[{{Komossa} {et~al.}(2017){Komossa}, {Grupe}, {Schartel}, {Gallo},
  {Gomez}, {Kollatschny}, {Kriss}, {Leighly}, {Longinotti}, {Parker},
  {Santos-Lleo}, {Wilkins}, \& {Zetzl}}]{ko17}
{Komossa}, S., {Grupe}, D., {Schartel}, N., {et~al.} 2017, in IAU Symposium,
  Vol. 324, New Frontiers in Black Hole Astrophysics, ed. A.~{Gomboc}, 168--171

\bibitem[{{Kova{\v{c}}evi{\'c}} {et~al.}(2018){Kova{\v{c}}evi{\'c}},
  {P{\'e}rez-Hern{\'a}ndez}, {Popovi{\'c}}, {Shapovalova}, {Kollatschny}, \&
  {Ili{\'c}}}]{ko18}
{Kova{\v{c}}evi{\'c}}, A.~B., {P{\'e}rez-Hern{\'a}ndez}, E., {Popovi{\'c}},
  L.~{\v{C}}., {et~al.} 2018, \mnras, 475, 2051

\bibitem[{{Kova{\v{c}}evi{\'c}} {et~al.}(2010){Kova{\v{c}}evi{\'c}},
  {Popovi{\'c}}, \& {Dimitrijevi{\'c}}}]{ko10}
{Kova{\v{c}}evi{\'c}}, J., {Popovi{\'c}}, L.~{\v{C}}., \& {Dimitrijevi{\'c}},
  M.~S. 2010, \apjs, 189, 15

\bibitem[{{Kraemer} {et~al.}(2002){Kraemer}, {Crenshaw}, {George}, {Netzer},
  {Turner}, \& {Gabel}}]{kr02}
{Kraemer}, S.~B., {Crenshaw}, D.~M., {George}, I.~M., {et~al.} 2002, \apj, 577,
  98

\bibitem[{{LaMassa} {et~al.}(2017){LaMassa}, {Yaqoob}, \& {Kilgard}}]{la17}
{LaMassa}, S.~M., {Yaqoob}, T., \& {Kilgard}, R. 2017, \apj, 840, 11

\bibitem[{{Landt} {et~al.}(2015a){Landt}, {Ward}, {Steenbrugge}, \&
  {Ferland}}]{la15a}
{Landt}, H., {Ward}, M.~J., {Steenbrugge}, K.~C., \& {Ferland}, G.~J. 2015a,
  \mnras, 449, 3795

\bibitem[{{Landt} {et~al.}(2015b){Landt}, {Ward}, {Steenbrugge}, \&
  {Ferland}}]{la15b}
{Landt}, H., {Ward}, M.~J., {Steenbrugge}, K.~C., \& {Ferland}, G.~J. 2015b,
  \mnras, 454, 3688

\bibitem[{{Liu} {et~al.}(2014){Liu}, {Wang}, {Yang}, {Zhu}, \& {Zhou}}]{li14}
{Liu}, T., {Wang}, J.-X., {Yang}, H., {Zhu}, F.-F., \& {Zhou}, Y.-Y. 2014,
  \apj, 783, 106

\bibitem[{{Lyuty} {et~al.}(1984){Lyuty}, {Oknyansky}, \& {Chuvaev}}]{ly84}
{Lyuty}, V.~M., {Oknyansky}, V.~L., \& {Chuvaev}, K.~K. 1984, Pisma v
  Astronomicheskii Zhurnal, 10, 803

\bibitem[{{Lyutyi}(1972)}]{ly72}
{Lyutyi}, V.~M. 1972, \azh, 49, 930 [Sov. Astron. 16, 763 (1972)]

\bibitem[{{Lyutyi} \& {Doroshenko}(1993)}]{ld93}
{Lyutyi}, V.~M. \& {Doroshenko}, V.~T. 1993, Astronomy Letters, 19, 405

\bibitem[{{MacLeod} {et~al.}(2019){MacLeod}, {Green}, {Anderson}, {Bruce},
  {Eracleous}, {Graham}, {Homan}, {Lawrence}, {LeBleu}, {Ross}, {Ruan},
  {Runnoe}, {Stern}, {Burgett}, {Chambers}, {Kaiser}, {Magnier}, \&
  {Metcalfe}}]{ml19}
{MacLeod}, C.~L., {Green}, P.~J., {Anderson}, S.~F., {et~al.} 2019, \apj, 874,
  8

\bibitem[{{MacLeod} {et~al.}(2016){MacLeod}, {Ross}, {Lawrence}, {Goad},
  {Horne}, {Burgett}, {Chambers}, {Flewelling}, {Hodapp}, {Kaiser}, {Magnier},
  {Wainscoat}, \& {Waters}}]{ml16}
{MacLeod}, C.~L., {Ross}, N.~P., {Lawrence}, A., {et~al.} 2016, \mnras, 457,
  389

\bibitem[{{Maoz} {et~al.}(2002){Maoz}, {Markowitz}, {Edelson}, \& {Nand
  ra}}]{ma02}
{Maoz}, D., {Markowitz}, A., {Edelson}, R., \& {Nand ra}, K. 2002, \aj, 124,
  1988

\bibitem[{{Mathur} {et~al.}(2018){Mathur}, {Denney}, {Gupta}, {Vestergaard},
  {De Rosa}, {Krongold}, {Nicastro}, {Collinson}, {Goad}, {Korista}, {Pogge},
  \& {Peterson}}]{ma18}
{Mathur}, S., {Denney}, K.~D., {Gupta}, A., {et~al.} 2018, \apj, 866, 123

\bibitem[{{Mehdipour} {et~al.}(2010){Mehdipour}, {Branduardi-Raymont}, \&
  {Page}}]{me10}
{Mehdipour}, M., {Branduardi-Raymont}, G., \& {Page}, M.~J. 2010, \aap, 514,
  A100

\bibitem[{{Netzer} {et~al.}(2002){Netzer}, {Chelouche}, {George}, {Turner},
  {Crenshaw}, {Kraemer}, \& {Nand ra}}]{ne02}
{Netzer}, H., {Chelouche}, D., {George}, I.~M., {et~al.} 2002, \apj, 571, 256

\bibitem[{{Noda} \& {Done}(2018)}]{nd18}
{Noda}, H. \& {Done}, C. 2018, \mnras, 480, 3898

\bibitem[{{Noda} {et~al.}(2016){Noda}, {Minezaki}, {Watanabe}, {Kokubo},
  {Kawaguchi}, {Itoh}, {Morihana}, {Saito}, {Nakao}, {Imai}, {Moritani},
  {Takaki}, {Kawabata}, {Nakaoka}, {Uemura}, {Kawabata}, {Yoshida}, {Arai},
  {Takagi}, {Morokuma}, {Doi}, {Itoh}, {Yamada}, {Nakazawa}, {Fukazawa}, \&
  {Makishima}}]{no16}
{Noda}, H., {Minezaki}, T., {Watanabe}, M., {et~al.} 2016, \apj, 828, 78

\bibitem[{{Oknyanskii} {et~al.}(1991){Oknyanskii}, {Lyutyi}, \&
  {Chuvaev}}]{ok91}
{Oknyanskii}, V.~L., {Lyutyi}, V.~M., \& {Chuvaev}, K.~K. 1991, Soviet
  Astronomy Letters, 17, 100

\bibitem[{{Oknyansky} {et~al.}(2017a){Oknyansky}, {Gaskell}, {Huseynov},
  {Lipunov}, {Shatsky}, {Tsygankov}, {Gorbovskoy}, {Mikailov}, {Tatarnikov},
  {Buckley}, {Metlov}, {Nadzhip}, {Kuznetsov}, {Balanutza}, {Burlak},
  {Galazutdinov}, {Artamonov}, {Salmanov}, {Malanchev}, \& {Oknyansky}}]{ok17a}
{Oknyansky}, V.~L., {Gaskell}, C.~M., {Huseynov}, N.~A., {et~al.} 2017a,
  \mnras, 467, 1496

\bibitem[{{Oknyansky} {et~al.}(2019a){Oknyansky}, {Winkler}, {Tsygankov},
  {Lipunov}, {Gorbovskoy}, {van Wyk}, {Buckley}, \& {Tyurina}}]{ok19a}
{Oknyansky}, V.~L., {Winkler}, H., {Tsygankov}, S.~S., {et~al.} 2019a, Odessa
  Astronomical Publications, 32, 75

\bibitem[{{Oknyansky} {et~al.}(2019b){Oknyansky}, {Winkler}, {Tsygankov},
  {Lipunov}, {Gorbovskoy}, {van Wyk}, {Buckley}, \& {Tyurina}}]{ok19b}
{Oknyansky}, V.~L., {Winkler}, H., {Tsygankov}, S.~S., {et~al.} 2019b, \mnras,
  483, 558

\bibitem[{{Peterson}(1988)}]{pe88}
{Peterson}, B.~M. 1988, \pasp, 100, 18

\bibitem[{{Pier} \& {Voit}(1995)}]{pv95}
{Pier}, E.~A. \& {Voit}, G.~M. 1995, in Oxford Torus Workshop, ed. M.~J.
  {Ward}, 93

\bibitem[{{Popovi{\'c}} {et~al.}(2004){Popovi{\'c}}, {Mediavilla}, {Bon}, \&
  {Ili{\'c}}}]{po04}
{Popovi{\'c}}, L.~{\v{C}}., {Mediavilla}, E., {Bon}, E., \& {Ili{\'c}}, D.
  2004, \aap, 423, 909

\bibitem[{{Popovi{\'c}} {et~al.}(2002){Popovi{\'c}}, {Mediavilla},
  {Kubi{\v{c}}ela}, \& {Jovanovi{\'c}}}]{po02}
{Popovi{\'c}}, L.~{\v{C}}., {Mediavilla}, E.~G., {Kubi{\v{c}}ela}, A., \&
  {Jovanovi{\'c}}, P. 2002, \aap, 390, 473

\bibitem[{{Popovi{\'c}} {et~al.}(2014){Popovi{\'c}}, {Shapovalova}, {Ili{\'c}},
  {Burenkov}, {Chavushyan}, {Kollatschny}, {Kova{\v{c}}evi{\'c}}, {Vald{\'e}s},
  {Le{\'o}n-Tavares}, {Bochkarev}, {Pati{\~n}o-{\'A}lvarez}, \&
  {Torrealba}}]{po14}
{Popovi{\'c}}, L.~{\v{C}}., {Shapovalova}, A.~I., {Ili{\'c}}, D., {et~al.}
  2014, \aap, 572, A66

\bibitem[{{Raki\'c}(2020)}]{rakic}
{Raki\'c}, N. 2020, in preparation

\bibitem[{{Rose} {et~al.}(2015){Rose}, {Elvis}, \& {Tadhunter}}]{ro15}
{Rose}, M., {Elvis}, M., \& {Tadhunter}, C.~N. 2015, \mnras, 448, 2900

\bibitem[{{Rumbaugh} {et~al.}(2018){Rumbaugh}, {Shen}, {Morganson}, {Liu},
  {Banerji}, {McMahon}, {Abdalla}, {Benoit-L{\'e}vy}, {Bertin}, {Brooks},
  {Buckley-Geer}, {Capozzi}, {Carnero Rosell}, {Carrasco Kind}, {Carretero},
  {Cunha}, {D'Andrea}, {da Costa}, {DePoy}, {Desai}, {Doel}, {Frieman},
  {Garc{\'\i}a-Bellido}, {Gruen}, {Gruendl}, {Gschwend}, {Gutierrez},
  {Honscheid}, {James}, {Kuehn}, {Kuhlmann}, {Kuropatkin}, {Lima}, {Maia},
  {Marshall}, {Martini}, {Menanteau}, {Plazas}, {Reil}, {Roodman}, {Sanchez},
  {Scarpine}, {Schindler}, {Schubnell}, {Sheldon}, {Smith}, {Soares-Santos},
  {Sobreira}, {Suchyta}, {Swanson}, {Walker}, {Wester}, \& {DES
  Collaboration}}]{ru18}
{Rumbaugh}, N., {Shen}, Y., {Morganson}, E., {et~al.} 2018, \apj, 854, 160

\bibitem[{{Runco} {et~al.}(2016){Runco}, {Cosens}, {Bennert}, {Scott},
  {Komossa}, {Malkan}, {Lazarova}, {Auger}, {Treu}, \& {Park}}]{ru16}
{Runco}, J.~N., {Cosens}, M., {Bennert}, V.~N., {et~al.} 2016, \apj, 821, 33

\bibitem[{{Sesar} {et~al.}(2007){Sesar}, {Ivezi{\'c}}, {Lupton}, {Juri{\'c}},
  {Gunn}, {Knapp}, {DeLee}, {Smith}, {Miknaitis}, {Lin}, {Tucker}, {Doi},
  {Tanaka}, {Fukugita}, {Holtzman}, {Kent}, {Yanny}, {Schlegel}, {Finkbeiner},
  {Padmanabhan}, {Rockosi}, {Bond}, {Lee}, {Stoughton}, {Jester}, {Harris},
  {Harding}, {Brinkmann}, {Schneider}, {York}, {Richmond}, \& {Vanden
  Berk}}]{se07}
{Sesar}, B., {Ivezi{\'c}}, {\v{Z}}., {Lupton}, R.~H., {et~al.} 2007, \aj, 134,
  2236

\bibitem[{{Seyfert}(1943)}]{se43}
{Seyfert}, C.~K. 1943, \apj, 97, 28

\bibitem[{{Shapovalova} {et~al.}(2019){Shapovalova}, {Popovi{\'c}}, {},
  {Afanasiev}, {Ili{\'c}}, {}, {Kova{\v{c}}evi{\'c}}, {}, {Burenkov},
  {Chavushyan}, {Mar{\v{c}}eta-Mandi{\'c}}, {}, {Spiridonova}, {Valdes},
  {Bochkarev}, {Pati{\~n}o-{\'A}lvarez}, {Carrasco}, \& {Zhdanova}}]{sh19}
{Shapovalova}, A.~I., {Popovi{\'c}}, {}, L.~{\v{C}}., {et~al.} 2019, \mnras,
  485, 4790

\bibitem[{{Shapovalova} {et~al.}(2012){Shapovalova}, {Popovi{\'c}}, {Burenkov},
  {Chavushyan}, {Ili{\'c}}, {Kova{\v{c}}evi{\'c}}, {Kollatschny},
  {Kova{\v{c}}evi{\'c}}, {Bochkarev}, {Valdes}, {Torrealba},
  {Le{\'o}n-Tavares}, {Mercado}, {Ben{\'\i}tez}, {Carrasco}, {Dultzin}, \& {de
  la Fuente}}]{sh12}
{Shapovalova}, A.~I., {Popovi{\'c}}, L.~{\v{C}}., {Burenkov}, A.~N., {et~al.}
  2012, \apjs, 202, 10

\bibitem[{{{\'S}niegowska} \& {Czerny}(2019)}]{sc19}
{{\'S}niegowska}, M. \& {Czerny}, B. 2019, arXiv e-prints, arXiv:1904.06767

\bibitem[{{Stern} {et~al.}(2018){Stern}, {McKernan}, {Graham}, {Ford}, {Ross},
  {Meisner}, {Assef}, {Balokovi{\'c}}, {Brightman}, {Dey}, {Drake},
  {Djorgovski}, {Eisenhardt}, \& {Jun}}]{st18}
{Stern}, D., {McKernan}, B., {Graham}, M.~J., {et~al.} 2018, \apj, 864, 27

\bibitem[{{Storchi-Bergmann} {et~al.}(2017){Storchi-Bergmann}, {Schimoia},
  {Peterson}, {Elvis}, {Denney}, {Eracleous}, \& {Nemmen}}]{sb17}
{Storchi-Bergmann}, T., {Schimoia}, J.~S., {Peterson}, B.~M., {et~al.} 2017,
  \apj, 835, 236

\bibitem[{{Turner} {et~al.}(2005){Turner}, {Kraemer}, {George}, {Reeves}, \&
  {Bottorff}}]{tu05}
{Turner}, T.~J., {Kraemer}, S.~B., {George}, I.~M., {Reeves}, J.~N., \&
  {Bottorff}, M.~C. 2005, \apj, 618, 155

\bibitem[{{Turner} {et~al.}(2011){Turner}, {Miller}, {Kraemer}, \&
  {Reeves}}]{tu11}
{Turner}, T.~J., {Miller}, L., {Kraemer}, S.~B., \& {Reeves}, J.~N. 2011, \apj,
  733, 48

\bibitem[{{Vanden Berk} {et~al.}(2006){Vanden Berk}, {Shen}, {Yip},
  {Schneider}, {Connolly}, {Burton}, {Jester}, {Hall}, {Szalay}, \&
  {Brinkmann}}]{vb06}
{Vanden Berk}, D.~E., {Shen}, J., {Yip}, C.-W., {et~al.} 2006, \aj, 131, 84

\bibitem[{{Vanden Berk} {et~al.}(2004){Vanden Berk}, {Wilhite}, {Kron},
  {Anderson}, {Brunner}, {Hall}, {Ivezi{\'c}}, {Richards}, {Schneider}, {York},
  {Brinkmann}, {Lamb}, {Nichol}, \& {Schlegel}}]{vb04}
{Vanden Berk}, D.~E., {Wilhite}, B.~C., {Kron}, R.~G., {et~al.} 2004, \apj,
  601, 692

\bibitem[{{Wachter} {et~al.}(1979){Wachter}, {Leach}, \& {Kellogg}}]{wa79}
{Wachter}, K., {Leach}, R., \& {Kellogg}, E. 1979, \apj, 230, 274

\bibitem[{{Wanders} \& {Horne}(1994)}]{wh94}
{Wanders}, I. \& {Horne}, K. 1994, \aap, 289, 76

\bibitem[{{Wanders} {et~al.}(1993){Wanders}, {van Groningen}, {Alloin},
  {Aretxaga}, {Axon}, {de Bruyn}, {Clavel}, {Dietrich}, {Goad}, {Gondhalekar},
  {Horne}, {Jackson}, {Kollatschny}, {Laurikainen}, {Lawrence}, {Masegosa},
  {O'Brien}, {del Olmo}, {Penston}, {Perea}, {Perez}, {Perez-Fournon}, {Perry},
  {Robinson}, {Rodriguez Espinosa}, {Stirpe}, {Tadhunter}, {Terlevich},
  {Unger}, {Wagner}, \& {Williams}}]{wa93}
{Wanders}, I., {van Groningen}, E., {Alloin}, D., {et~al.} 1993, \aap, 269, 39

\bibitem[{{Wang} {et~al.}(2012){Wang}, {Zhou}, {Komossa}, {Wang}, {Yuan}, \&
  {Yang}}]{wa12}
{Wang}, T.-G., {Zhou}, H.-Y., {Komossa}, S., {et~al.} 2012, \apj, 749, 115

\bibitem[{{Yan} {et~al.}(2019){Yan}, {Wang}, {Jiang}, {Stern}, {Dou},
  {Fremling}, {Graham}, {Drake}, {Yang}, {Burdge}, \& {Kasliwal}}]{ya19}
{Yan}, L., {Wang}, T., {Jiang}, N., {et~al.} 2019, \apj, 874, 44

\bibitem[{{Yang} {et~al.}(2013){Yang}, {Wang}, {Ferland}, {Yuan}, {Zhou}, \&
  {Jiang}}]{ya13}
{Yang}, C.-W., {Wang}, T.-G., {Ferland}, G., {et~al.} 2013, \apj, 774, 46

\end{thebibliography}

\end{document}